# Fracturing of polycrystalline MoS$_2$ nanofilms

M. Sledzinska[1*], G. Jumbert[1,2], M. Placidi[3], A. Arrighi[1,2], P. Xiao[1,2], F. Alzina[1] and C.M. Sotomayor Torres[1,4]

[1] Catalan Institute of Nanoscience and Nanotechnology (ICN2), CSIC and BIST, Campus UAB, Bellaterra, 08193 Barcelona, Spain
[2] Departament de Fisica, Universitat Autoonoma de Barcelona, Bellaterra, E-08193 Barcelona, Spain
[3] Catalonia Institute for Energy Research (IREC), Jardíns de les Dones de Negre 1, E-08930, Sant Adrià de Besòs, Spain
[4] ICREA, Pg. Lluís Companys 23, 08010 Barcelona, Spain

*Corresponding author: marianna.sledzinska@icn2.cat

**Abstract**

The possibility of tailoring the critical strain of 2D materials will be crucial for the fabrication of flexible devices. In this paper, the fracture in polycrystalline MoS$_2$ films with two different grain orientations is studied at the micro- and nanoscale using electron microscopy. The critical uniaxial strain is determined to be approximately 5% and independent of the sample morphology. However, electron beam irradiation is found to enhance the interaction between the MoS$_2$ and the PDMS substrates, leading to an increased critical strain that can exceed 10%. This enhancement of strain resistance was used to fabricate a mechanically robust array of lines 1 mm in length. Finally, nanoscale crack propagation studied by transmission electron microscopy showed that cracks propagate along the grain boundaries as well as through the grains, preferentially along van der Waals bonding. These results provide insight regarding the fracture of polycrystalline 2D materials and a new method to tailor the critical strain and nanofabrication of ultra-thin MoS$_2$ devices using well-developed tools, which will be of great interest to the flexible electronics industry.

**Introduction**

Recent years have seen a rapid increase in demand for personal flexible and wearable devices [1]. With the development of 2D materials, the pursuit of high-performance flexible devices would seem even closer to realisation. Two-dimensional materials are potential candidates for flexible nanotechnology because of their unique physical and mechanical properties and their ease of transfer onto soft and plastic substrates that facilitates large-area flexible devices fabricated at a reasonable cost [2]. One of the key requirements for flexible electronics is the ability of the active components to handle a reasonably high strain. Therefore, it has been crucial to exploit the properties of 2D materials to obtain properly functioning flexible devices. Of the myriad 2D materials, graphene is considered the strongest material ever measured. It possesses a Young's modulus of 1.0 TPa and an intrinsic strength of 130 GPa, and has been shown to handle deformations beyond the linear regime [3]. In the case of transition metal dichalcogenides, a single-crystal, single-layer $MoS_2$ film was shown to possess a Young's modulus of about 270 GPa [4,5], and breaking of the film was reported to occur at an effective strain between 6 and 11% for $MoS_2$ single and bi-layers, respectively.

Though single crystals of 2D materials possess outstanding elastic properties, point and line defects within these materials will play an important role in potential technological applications of large-area samples grown by methods such as chemical vapour deposition (CVD). In the work of Lee et al. [6], the mechanical properties of polycrystalline graphene films were studied by means of transmission electron microscopy (TEM) with nanoindentation. It was shown that, if the proper fabrication steps are followed, the stiffness of CVD-graphene is identical to that of single-crystalline graphene and the grain boundaries are almost as strong as those in pristine graphene.

Nevertheless, a systematic experimental and theoretical investigation of the deteriorating effects of structural of defects is lacking. The general trend in polycrystalline materials is a decrease of the Young's modulus, strength, toughness, and strain energy release rate and an increase of the critical strain with a decrease of a grain size [7-9]. In particular, the polycrystalline structure of $MoS_2$ makes it mechanically soft. Previously, we reported the Young's modulus of 20 (16) GPa for 5 (10) nm-thick $MoS_2$ membranes, which is a significant reduction (i.e., approximately five- to eight-fold) with respect to the bulk crystal [10]. Furthermore, the Young's modulus of the 5 and 10 nm-thick $MoS_2$ membranes is 16 times smaller than that of single-crystal few-layer $MoS_2$ membranes [4].

In this work, we further study the elastic properties of polycrystalline $MoS_2$ films in the limit of very small grains and in two different grain morphologies. We investigate the critical strain that this material can endure before breaking when placed on different stretchable substrates, such as elastic tape and PDMS by scanning electron microscopy (SEM). Moreover, we investigate the effect that electron beam exposure has on the $MoS_2$/PDMS interface, as a means to increase the critical strain. Finally, we study the morphology and propagation of nano-cracks forming in the $MoS_2$ by TEM. The aim of this work is to explore strategies to maximise the strain that the material can endure without breaking and identify optimised materials that can endure the unprecedented stress levels of 2D materials.

**Results**

*1. Measurements of critical strain in MoS$_2$*

Herein, we consider two types of MoS$_2$ with different in grain orientation with respect to the film plane, namely a 17 nm-thick film containing both horizontally- and vertically-oriented grains (H-V-type) and a 3 nm-thick film containing only horizontally-oriented grains (H-type) (See Supplementary Information, Fig. S1 and S2 for sample description).

To study the critical strain in MoS$_2$ we designed and built a tool for uniaxial stretching of the samples in a controlled manner inside the SEM chamber, as shown schematically in Fig. 1(a). The device was based on a micrometric screw that varied the distance between two clamps to which the elastic substrates were attached. The change in the substrate length was measured using a digital calliper and SEM image analysis.

Figure 1(b) shows an SEM image of the initial H-V-type film on elastic tape, where the sample already exhibits some cracks that occurred during wet transfer. From the measurements in the SEM we can estimate the critical tensile strain that can be endured by the MoS$_2$ film before micro-cracks appear in the material. For example, the initial MoS$_2$ film (Fig. 1(b)) can be compared with a film under 4.5% strain (Fig. 1(c)), where the latter shows fresh micro-fractures that are clearly visible and remain visible even when the stress is released. (Fig. 1(d)).

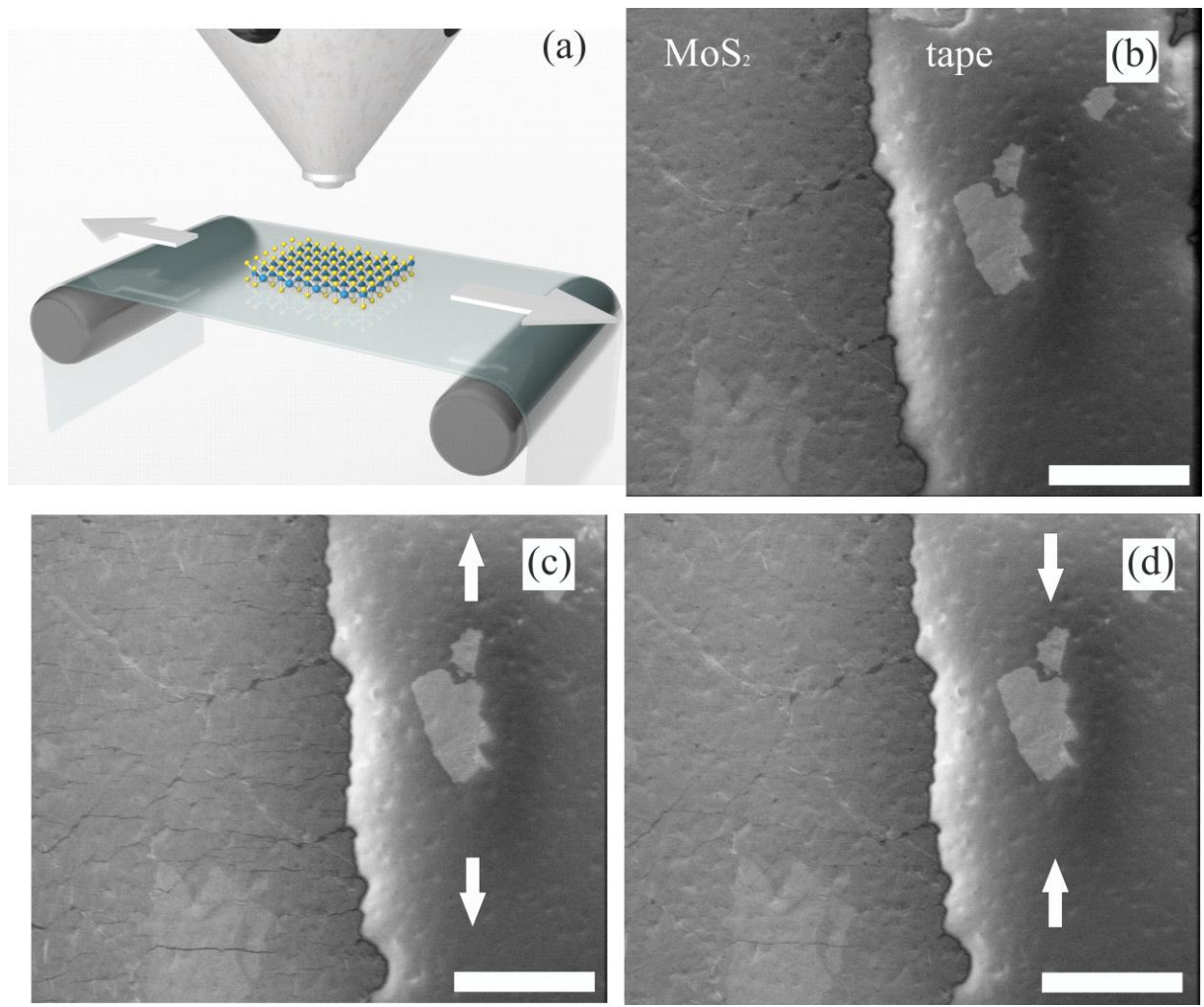

**Figure 1**. Fracturing of polycrystalline MoS$_2$ films on elastic tape. (a) Schematic of the experimental setup for critical strain monitoring inside the SEM chamber. SEM images of the (b) initial sample, 0% strain; (c) 4.5% strain; and (d) released, 0% strain. Arrows in (c) and (d) indicate stretching direction. Scale bars correspond to 50 μm.

Critical tensile strain of 5.6 ± 2% was measured for the H-V-type MoS$_2$ film on an elastic tape and on PDMS substrates in four independent measurements. A comparable critical strain value of 4.9 ± 2% was found also for the H-type film (Table S1 in Supplementary Information). Using the Young's moduli reported previously for these samples, we obtain breaking strength values of 0.9 and 1 GPa for the H-V- and H-type films, respectively. These polycrystalline films possess much lower breaking strength values than the reported single-crystalline film value of 23 GPa [4]. We attribute this behaviour to the small grain sizes in the polycrystalline films and the resulting high volume fraction of atoms constituting grain boundaries. We also note that there is no significant difference between the breaking strengths of these two types of samples i.e., H-V- and H-type films.

During the critical strain measurements, after each stretching iteration we obtained images of the region of interest at three magnifications (500×, 1500× and 2500×). Figures 2(a) and 2(b) show an example of a $MoS_2$/PDMS-mounted film before and after stretching, respectively. It can be seen in Figs. 2(a) and 2(b), and we noted the effect in all of the PDMS-mounted samples, that micro-cracks appeared in the $MoS_2$ film after exceeding the critical strain, but only in the regions where the high-resolution images were not obtained (area outside white box in Fig. 2(b)). On the contrary, in the regions where high-resolution (2500×) images were obtained (inside the white box in Fig. 2(b)), no cracks were visible. We further increased the strain and observed no pronounced cracks appearing in these electron-beam-exposed regions, up to approximately 20% strain. We further examined the sample in a high-resolution SEM (Fig. 1(c)), and the upper region exposed to a high electron dose from the electron beam presents no visible cracks (Fig. 2(d)), while micro-cracks are clearly visible in the unexposed lower region (Fig. 2(e)). This behaviour was observed for both H-V- and H-type $MoS_2$ films, but only when mounted on the PDMS substrates.

It has been reported that exposing PDMS to an electron beam modifies the material through electrostatic mechanisms[11] and induces cross-linking of the elastomer[12]. Monte Carlo simulations presented in Ref. [12] indicate that most of the electron energy is deposited in the top part of the PDMS film surface; therefore the interface between the $MoS_2$ film and the PDMS experienced the strongest modification. This effect happened at higher magnifications, where the effective electron dose per area was higher. One of the possible results of this modification could be a stronger adhesion of the $MoS_2$ to the PDMS substrate, enhanced by the short-bonded PDMS chains. It has been also reported that electron-beam exposure induces a change of the elastic properties of PDMS, such as an increased Young's modulus from ~50 to ~350 MPa [12]. However, no chemical modification of the PDMS surface has been previously detected as a result of electron beam exposure [11]. Also in this work, no chemical modification of the exposed $MoS_2$ and PDMS was detected using Raman spectroscopy (see Fig S3 and S4).

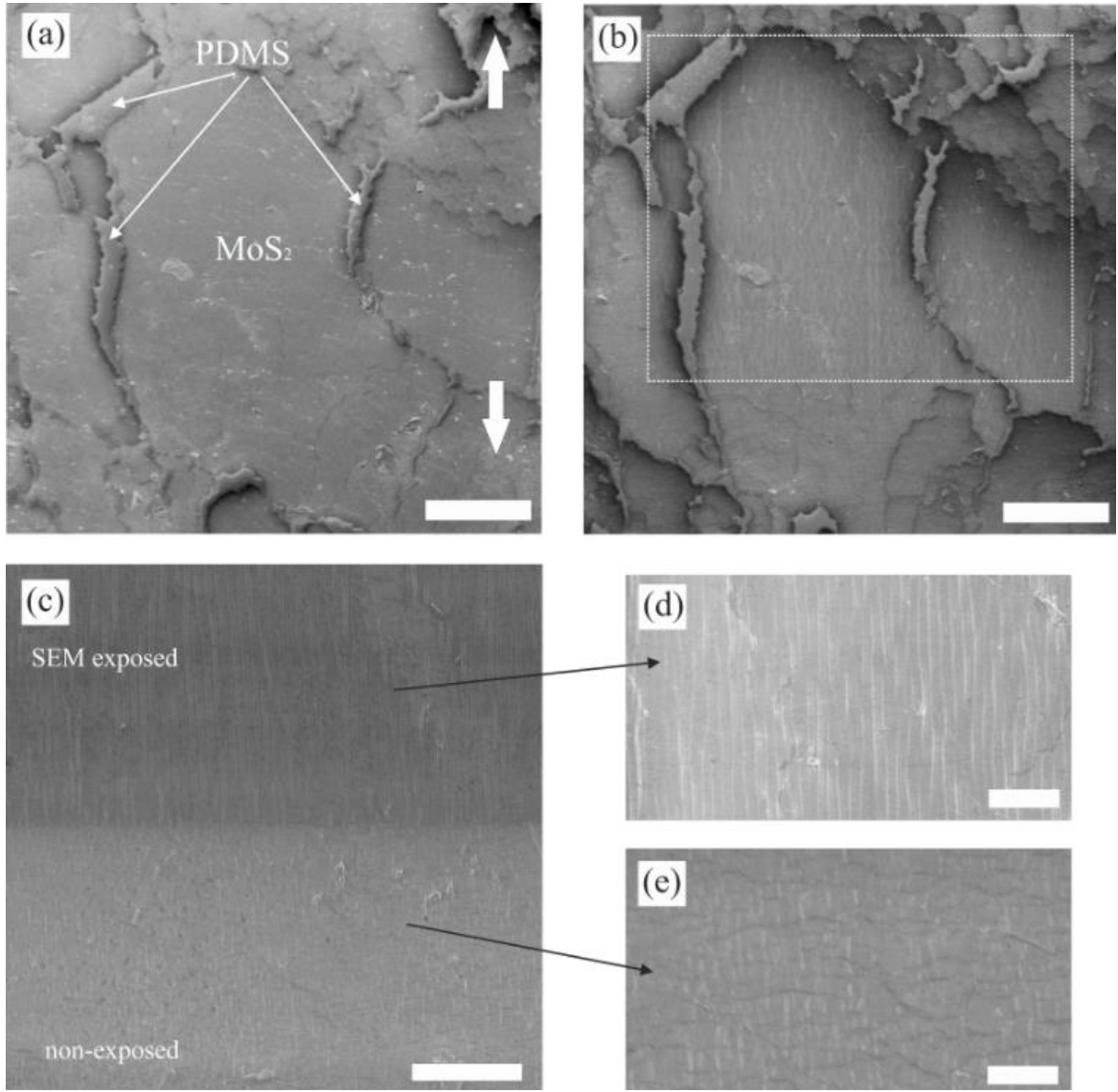

**Figure 2.** Fracturing of polycrystalline MoS$_2$ films on PDMS. SEM images of (a) initial sample, 0% strain; (b) stretched sample, 17% strain. White box indicates the area where high-magnification (2500×) SEM images were taken and the arrows indicate the stretching direction. Scale bars correspond to 100 µm. (c) Higher resolution SEM image of the boundary between the SEM-exposed region (upper) and unexposed region (lower). Scale bar 25 µm. High-resolution images of the (d) SEM-exposed area with no micro-cracks visible and (e) the unexposed area with micro-cracks visible. Scale bars 10 µm.

## 2. Tailoring of the critical strain in MoS$_2$ using electron beam lithography

The possibility of tailoring critical strain of the material by modifying its interactions with the substrate can be useful for the flexible device industry. We validated this concept further by performing electron-beam lithography (EBL) on the MoS$_2$/PDMS samples. The developed proof-of-concept sample comprised EBL of an array of 1 mm-long lines 100 µm apart with line widths ranging from 2.5 to 20 µm, where the electron beam dose was varied between 600 and 2000 µC/cm$^2$. An optical contrast between the exposed and unexposed areas is clearly seen under the optical microscope, confirming the cross-linking which occurs in the PDMS substrate after electron exposure (Fig. 3(a)). Further, the same area possessed structures visible using SEM (Fig. 3(b)). We note that these exposure-induced structures are well-defined only in the regions where MoS$_2$ is placed on the PDMS (see Fig S5). In this case the MoS$_2$ acts as a conductive layer in a similar way in which conductive polymers are used for EBL to avoid charging of insulating substrates.

The MoS$_2$ film is considerably wrinkled around the exposed structures, while the exposed areas appear smooth (Fig. 3(c)). The EBL-exposed sample, stretched above the critical strain (>10%), was imaged in the SEM and under the optical microscope. The unexposed area exhibited numerous fractures while the exposed area remained uniform (Figs. 3(c), 3(d) and S6). Therefore, we prove that electron-beam exposure can locally enhance the strain resistance of the MoS$_2$/PDMS to values above the critical limit, where the latter value was measured in the previous sections of this work. Using EBL, it is possible to write structures with a resolution of approximately 1 µm (Fig. S7).

The samples were also studied using atomic force microscopy (AFM). Figure 3(e) shows an AFM image of the MoS$_2$/PDMS sample, where it can be seen that the electron-beam exposed PDMS contracts. Similar, dose-dependent elastomer contraction was reported previously[12]. This result explains the appearance of wrinkles in the MoS$_2$ film around the exposed regions. The depth of the channels produced by the lines of contracted PDMS varies proportionally with the exposure dose. The lowest and highest doses (600 and 2000 µC/cm$^2$) resulted in 104 and 350 nm trench depth, respectively (see Table S2).

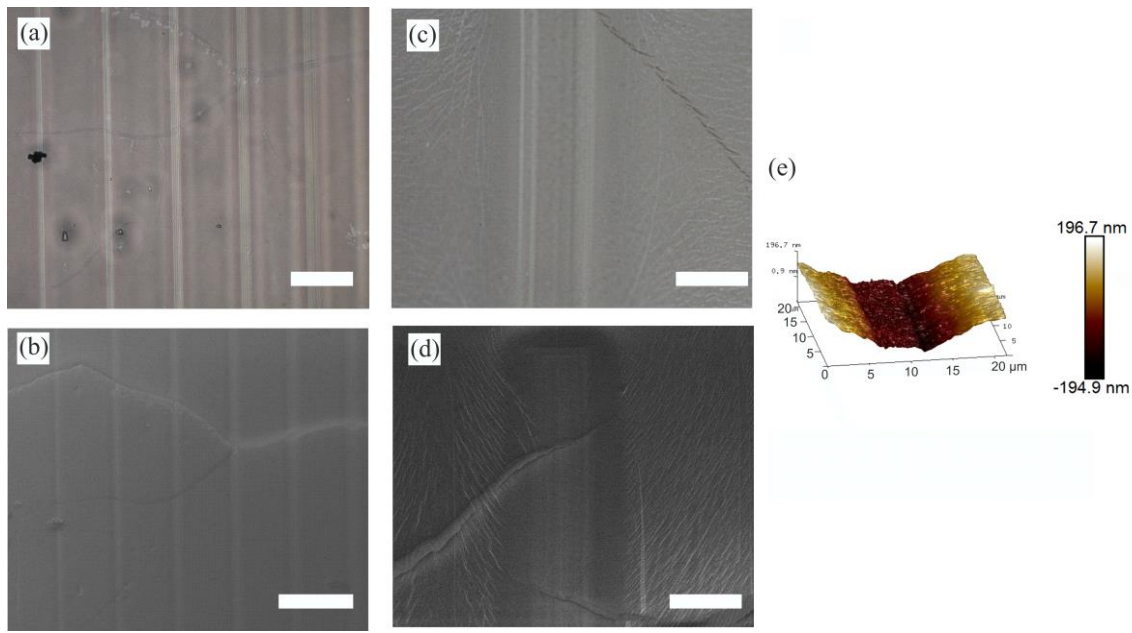

**Figure 3**. Electron beam lithography on MoS$_2$/PDMS samples. Sample area with 1 mm-long lines exposed with an electron dose of 1000 µC/cm$^2$ imaged by (a) optical microscope and (b) SEM. Scale bars correspond to 100 µm. Single line after >10% strain was applied on the sample imaged by (c) optical microscope and (d) SEM. Scale bars correspond to 25 µm. (e) Atomic force micrograph of the 5 µm-wide line.

*3. Nanoscale crack propagation in polycrystalline MoS$_2$*

Finally, we investigated the atomic structure of the cracked regions in the H-V- and H-type MoS$_2$ film samples using TEM. In case of the H-type films we were able to study crack propagation *in situ.* A pre-existing nano-crack (Figs. 4(a) and 4(b)) was exposed to the TEM electron beam. The crack propagated only due to the electron beam exposure, no external pulling force was applied. Multiple images of the propagation of the crack were recorded as a function of time, producing a video of the propagation (see SI). Figures 4(c)–4(j) are snapshots from the video. We first note that the fracture occurs at the grain boundaries, marked with arrows in Fig. 4(b). This is due to the fact that the covalent bonding within the grains is much stronger than the bonding between adjacent grains, and the position of the fracture confirms that the grain boundaries are the weakest points in the system.

Second, we note that the crack tip is not atomically sharp, but rather has a rounded form. Previous studies on crystalline MoS$_2$ have shown atomic sharpness at a crack tip without any blunting, which is a signature of pure brittleness and is expected for the covalent bonding of MoS$_2$ [13] [14]. Brittle fracture is associated with rapid clean cleavage of the bonds while ductile fracture generally indicates some plastic deformation around the tip front that slows crack propagation. In the case of polycrystalline MoS$_2$, the plastic deformation is clearly visible and is a signature of a ductile fracture.

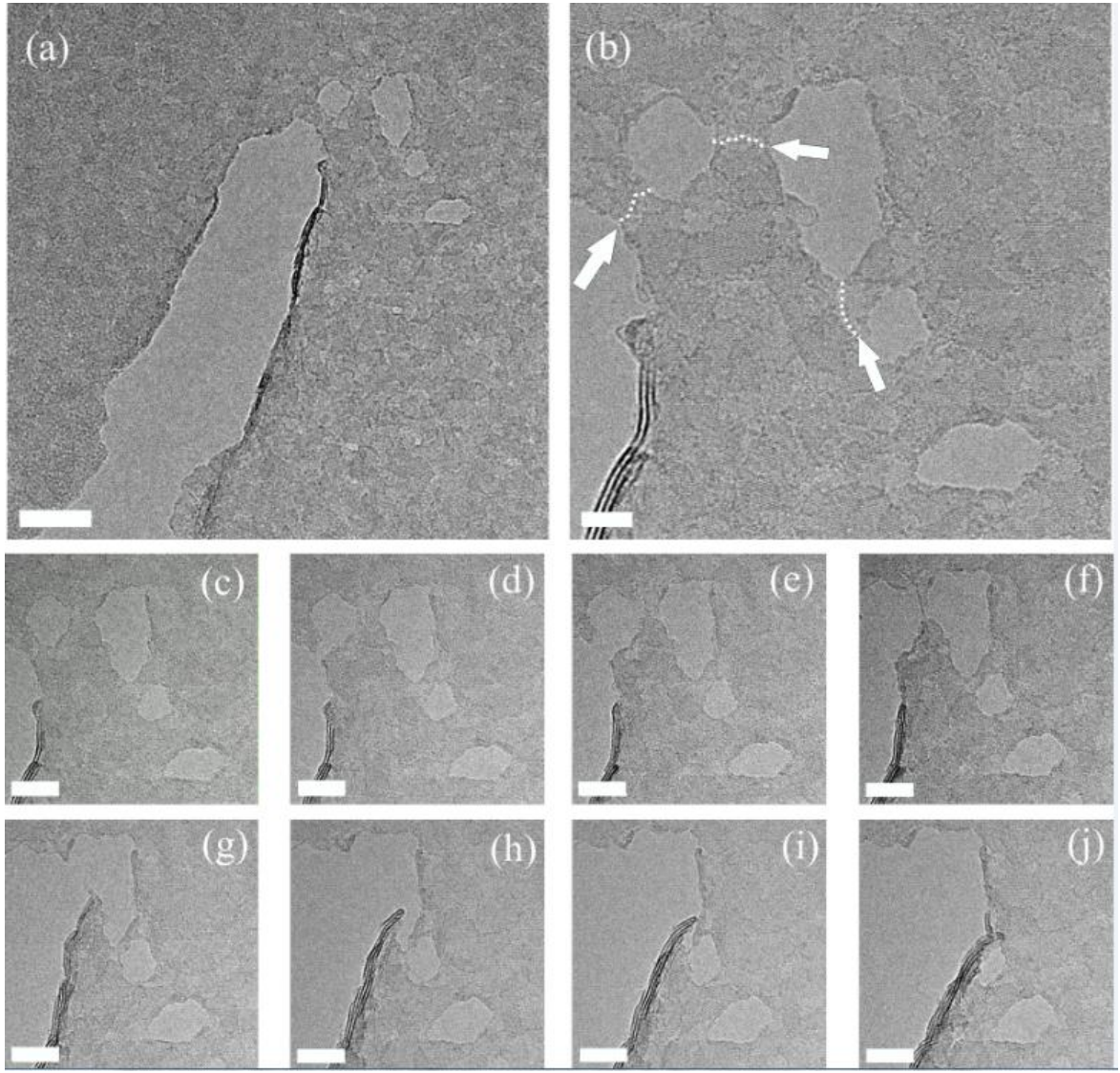

**Figure 4.** TEM images of the MoS$_2$ H-type film under the electron beam of the TEM. (a) Overview of the crack zone. Scale bar 20 nm. (b) Grain boundaries subject to fracture are marked with dashed lines and arrows. Scale bar 5 nm. (c–j) Images obtained as a function of time showing the crack propagation in MoS$_2$ film. Scale bars 10 nm.

Owing to the larger thickness of the H-V-type samples it was not possible to study crack formation *in situ*. Instead, for the H-V-type MoS$_2$ film we studied a nano-crack (approx. 10 nm wide) already present in the film. The width of the crack allowed us to observe both of its edges in the same image. Further, the presence of both vertically-oriented grains (VOGs) and horizontally-oriented grains (HOGs; see Fig S1) enabled the study of various types of interfaces. Figures 5(a) and 5(b) show an overview of the already-existing crack with a total length of more than 1 μm. We obtained high-resolution TEM images of various zones along the crack, whereby we could study the breaking pattern in the film. First, Figs. 5(c) and 5(d) show that the sample fractures on the grain boundaries in the case of two adjacent HOGs (marked with arrows 1 and 2 in Fig. 5(d)). This is the same result observed in the case of H-type films discussed above.

Second, the case of two adjacent grains with different orientations (Figs. 5(e) and 5(f)) shows that the film breaks at the interface between the HOG and VOG (indicated by arrow 1 in Fig. 5(f)). We have discussed this interface in our previous work and have postulated that there must be a considerable amount of amorphous phase between two grains of different orientations[15]. In this experiment we confirm that the bonding between the grains with different orientations is very weak.

Finally, in the case of adjacent VOGs (indicated by arrow 2 in Fig. 5(f)), rupture can occur both on the grain boundaries as well as inside the grains. This can be explained by the fact that the atomic layers within the grain and the grain boundaries are all bound by the weak van der Waals interaction, and thus there should be little distinction between the strength of the grains and the grain boundaries.

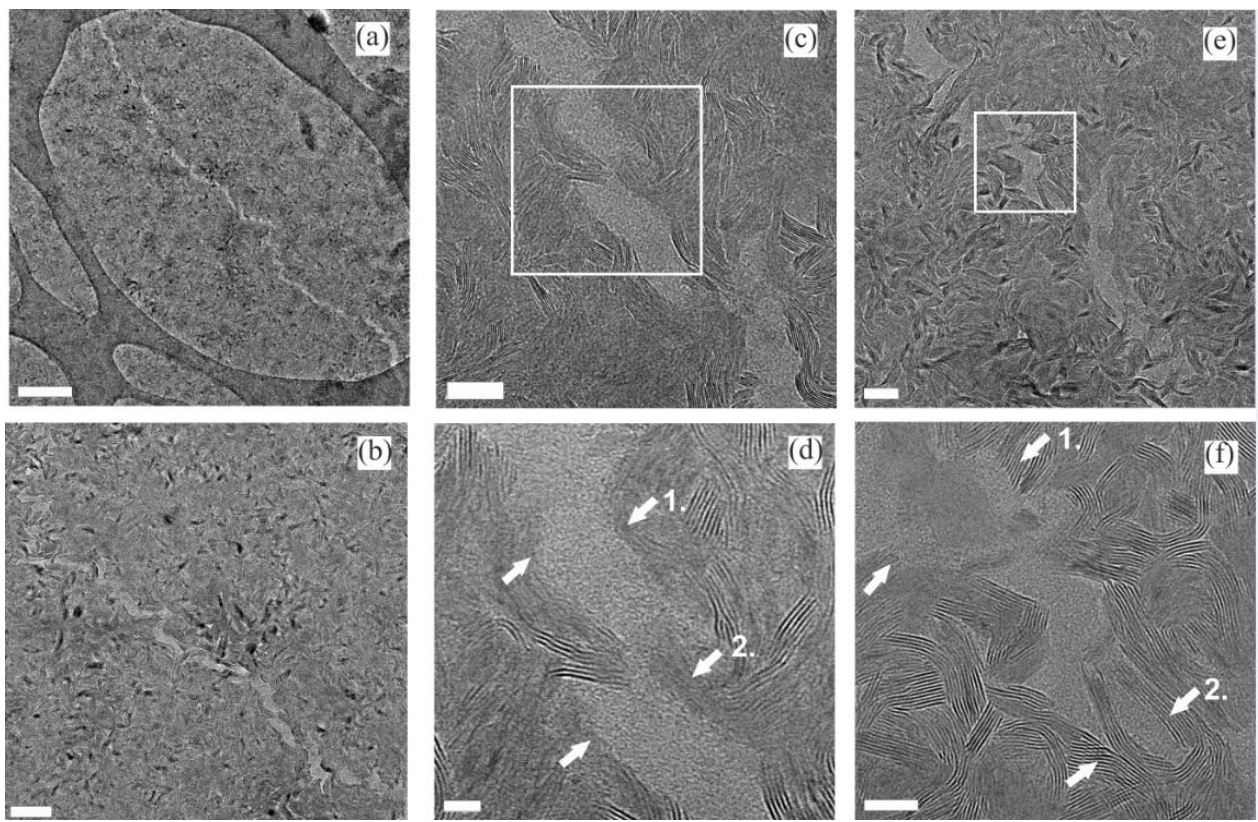

**Figure 5.** TEM images of crack propagation in H-V-type MoS$_2$ films. Overview of the nano-crack present in the sample, scale bars (a) 250 nm and (b) 50 nm. (c–f) Details of the nano-crack: scale bars (c) 10 nm, (d) 5 nm, (e) 20 nm, (f) 10 nm.

**Conclusions**

We measured the critical tensile strain of polycrystalline MoS$_2$ films, which is on the order of 5%, where the value was found to be independent of the sample thickness and morphology. This critical strain value is lower than those previously reported for crystalline single- and bi-layer MoS$_2$ samples, confirming the limiting influence of the grain boundaries. However, the critical strain value of the MoS$_2$ film supported on a PDMS substrate was increased by electron beam exposure, withstanding strains exceeding 10% with no micro-crack formation. As the cross-linking occurs in the PDMS substrate it should in principle be possible to extend this technique to other 2D materials.

We also studied the nanoscale fracture of H- and H-V-type $MoS_2$ films at the atomic level via TEM. We have confirmed that in polycrystalline $MoS_2$ cracks propagate mainly along the grain boundaries. Also the weak van der Waals forces between the $MoS_2$ layers within a grain and the at the interface between grains are the most prone to fracture, constituting the weakest links in the samples.

In the future, understanding of the role of crystalline structure of the 2D materials and its interaction with a flexible substrate may result in enhanced strain resistance for flexible devices.

**Experimental Methods**

*Sample fabrication*

In this work we studied $MoS_2$ with two different grain morphologies: horizontally- and vertically-oriented grains (H-V-type) and only horizontally-oriented grains (H-type).

The synthesis of the $MoS_2$ nanosheets was achieved via a thin-film conversion technique. A thin Mo layer with varying thickness was deposited by DC-magnetron sputtering on $10\times10$ cm$^2$ glass substrates and was reactively annealed in a graphite box in a sulphur-containing atmosphere at 600°C for 30 min [15-16].

Polymer-free, water-assisted transfer was used to place the $MoS_2$ films on the substrates used for testing; including TEM grids for TEM analysis and two flexible substrates of elastic tape (Nitto) and thin PDMS film (Gel Pak) for strain analysis.

*$MoS_2$ characterisation*

Atomic force microscope topographical images were obtained using a Nanoscope IV controller and a Dimension 3100 head (Veeco).

Critical strain experiments were performed in the FEI Quanta 650FEG ESEM at 2kV. High-resolution SEM images were obtained in the FEI MAGELLAN 400L.

Raman scattering measurements were performed using the Horiba T64000 Raman spectrometer and a 532 nm laser (Cobolt). The Raman spectra at each point was obtained at the same incident laser power (0.3 mW) and integrated for the same amount of time (180 s).

High-resolution imaging of the structure and morphology of the samples was obtained using the FEI Tecnai F20 in TEM and scanning TEM (STEM) modes.

Electron beam lithography was performed using the Inspect F50 FEI SEM-based system at 30 kV acceleration voltage and write fields of 1×1 mm and 200×200 µm.

**Acknowledgements**

ICN2 is funded by the CERCA program/Generalitat de Catalunya, and is supported by the Severo Ochoa program from Spanish MINECO (Grant No. SEV-2017-0706). We acknowledge support from EU Projects NANOSMART (H2020 ICT-07-2018) and NANOPOLY (H2020 FETOPEN-01-2018-2019-2020) and the Spanish MICINN project SIP (PGC2018-101743-B-I00).

**References**

1. *Market Research Report Flexible Electronics Market by Components (Display, Battery, Sensors, Memory), by Application (Consumer Electronics, Automotive, Healthcare, Industrial) and Segment Forecast to 2024.*; Grand View Research **2016**.
2. Akinwande, D.; Petrone, N.; Hone, J. Two-Dimensional Flexible Nanoelectronics. *Nature communications* **2014,** *5*, 5678.
3. Lee, C.; Wei, X.; Kysar, J. W.; Hone, J. Measurement of the Elastic Properties and Intrinsic Strength of Monolayer Graphene. *Science* **2008,** *321*, 385-8.
4. Bertolazzi, S.; Brivio, J.; Kis, A. Stretching and Breaking of Ultrathin Mos2. *ACS nano* **2011,** *5* 9703–9709.
5. Liu, K.; Yan, Q.; Chen, M.; Fan, W.; Sun, Y.; Suh, J.; Fu, D.; Lee, S.; Zhou, J.; Tongay, S.; Ji, J.; Neaton, J. B.; Wu, J. Elastic Properties of Chemical-Vapor-Deposited Monolayer Mos2, Ws2, and Their Bilayer Heterostructures. *Nano letters* **2014,** *14*, 5097-103.
6. Lee, G.-H.; Cooper, R. C.; An, S. J.; Lee, S.; van der Zande, A.; Petrone, N.; Hammerberg, A. G.; Lee, C.; Crawford, B.; Oliver, W.; Kysar, J. W.; Hone, J. High-Strength Chemical-Vapor–Deposited Graphene and Grain Boundaries. *Science* **2013,** *340*, 1073-1078.
7. Hess, P. Relationships between the Elastic and Fracture Properties of Boronitrene and Molybdenum Disulfide and Those of Graphene. *Nanotechnology* **2017,** *28*, 064002.
8. Becton, M.; Wang, X. Grain-Size Dependence of Mechanical Properties in Polycrystalline Boron-Nitride: A Computational Study. *Physical chemistry chemical physics : PCCP* **2015,** *17*, 21894-901.
9. Mortazavi, B.; Cuniberti, G. Mechanical Properties of Polycrystalline Boron-Nitride Nanosheets. *RSC Adv.* **2014,** *4*, 19137-19143.
10. Graczykowski, B.; Sledzinska, M.; Placidi, M.; Saleta Reig, D.; Kasprzak, M.; Alzina, F.; Sotomayor Torres, C. M. Elastic Properties of Few Nanometers Thick Polycrystalline Mos2 Membranes: A Nondestructive Study. *Nano letters* **2017,** *17*, 7647-7651.
11. Russell, M. T.; Pingree, L. S. C.; Hersam, M. C.; Marks, T. J. Microscale Features and Surface Chemical Functionality Patterned by Electron Beam Lithography: A Novel Route to Poly(Dimethylsiloxane) (Pdms) Stamp Fabrication. *Langmuir* **2006,** *22*, 6712-6718.
12. Biggs, M. J. P.; Fernandez, M.; Thomas, D.; Cooper, R.; Palma, M.; Liao, J.; Fazio, T.; Dahlberg, C.; Wheadon, H.; Pallipurath, A.; Pandit, A.; Kysar, J.; Wind, S. J. The Functional Response of Mesenchymal Stem Cells to Electron-Beam Patterned Elastomeric Surfaces Presenting Micrometer to Nanoscale Heterogeneous Rigidity. *Advanced materials* **2017,** *29*.
13. Wang, S.; Qin, Z.; Jung, G. S.; Martin-Martinez, F. J.; Zhang, K.; Buehler, M. J.; Warner, J. H. Atomically Sharp Crack Tips in Monolayer Mos2 and Their Enhanced Toughness by Vacancy Defects. *ACS nano* **2016,** *10*, 9831-9839.
14. Ly, T. H.; Zhao, J.; Cichocka, M. O.; Li, L. J.; Lee, Y. H. Dynamical Observations on the Crack Tip Zone and Stress Corrosion of Two-Dimensional Mos2. *Nature communications* **2017,** *8*, 14116.
15. Sledzinska, M.; Quey, R.; Mortazavi, B.; Graczykowski, B.; Placidi, M.; Saleta Reig, D.; Navarro-Urrios, D.; Alzina, F.; Colombo, L.; Roche, S.; Sotomayor Torres, C. M. Record Low

Thermal Conductivity of Polycrystalline Mos2 Films: Tuning the Thermal Conductivity by Grain Orientation. *ACS applied materials & interfaces* **2017,** *9*, 37905-37911.
16.     Sledzinska, M.; Graczykowski, B.; Placidi, M.; Reig, D. S.; Sachat, A. E.; Reparaz, J. S.; Alzina, F.; Mortazavi, B.; Quey, R.; Colombo, L.; Roche, S.; Torres, C. M. S. Thermal Conductivity of Mos2 Polycrystalline Nanomembranes. *2D Materials* **2016,** *3*, 035016-035022.